\documentclass[10pt, modern]{AASTexv611/aastex61}

\usepackage{amsmath,amsthm,amsfonts,latexsym,amscd, tabularx}
\usepackage{graphicx}
\usepackage{longtable}
\usepackage{array}
\usepackage{gensymb}
\usepackage[margin=1.0in]{geometry}
\usepackage{lipsum}
\usepackage{amsmath}
\usepackage{amssymb}
\usepackage{color,soul}

\newcommand{\kms}{\ensuremath{{\rm km~s^{-1}}}}
\newcommand{\hcop}{\text{HCO\textsuperscript{+}}}

\newcommand{\persc}{\ensuremath{\,{\rm cm^{-2}}}}
\newcommand{\htwo}{\text{H\textsubscript{2}}}

\graphicspath{ {Figures/} }

\begin{document}

\title{A Search for 3-mm Molecular Absorption Line Transitions in the Magellanic Stream}

\author{Lucille Steffes}
\affiliation{University of Wisconsin--Madison, Department of Astronomy, 475 N Charter St, Madison, WI 53703, USA}

\author{Daniel R. Rybarczyk}
\affiliation{University of Wisconsin--Madison, Department of Astronomy, 475 N Charter St, Madison, WI 53703, USA}

\author{Sne\v zana Stanimirovi\'c}
\affiliation{University of Wisconsin--Madison, Department of Astronomy, 475 N Charter St, Madison, WI 53703, USA}

\author{J. R. Dawson}
\affiliation{School of Mathematical and Physical Sciences and Research Centre in Astronomy, Astrophysics \& Astrophotonics, Macquarie University, Sydney 2109,
Australia}
\affiliation{Australia Telescope National Facility, CSIRO Space \& Astronomy, PO Box 76, Epping, NSW 1710, Australia}

\author{Mary Putman} 
\affiliation{Department of Astronomy, Columbia University, New York, NY 10027, USA}

\author{Philipp Richter}
\affiliation{Institut für Physik und Astronomie, Universität Potsdam, Haus 28, Karl-Liebknecht-Str. 24/25, D-14476 Golm (Potsdam), Germany}

\author{John Gallagher III} 
\affiliation{University of Wisconsin--Madison, Department of Astronomy, 475 N Charter St, Madison, WI 53703, USA}

\author{Harvey Liszt} 
\affiliation{National Radio Astronomy Observatory, 520 Edgemont Road, Charlottesville, VA 22903, USA}

\author{Claire Murray}
\affiliation{University of Wisconsin--Madison, Department of Astronomy, 475 N Charter St, Madison, WI 53703, USA}
\affiliation{current: Space Telescope Science Institute, 3700 San Martin Dr, Baltimore, MD 21218}

\author{John Dickey} 
\affiliation{School of Natural Sciences, University of Tasmania, Hobart, TAS 7001, Australia}

\author{Carl Heiles} 
\affiliation{Department of Astronomy, University of California, Berkeley, 601 Campbell Hall 3411, Berkeley, CA 94720-3411, USA}

\author{Audra Hernandez} 
\affiliation{University of Wisconsin--Madison, Department of Astronomy, 475 N Charter St, Madison, WI 53703, USA}

\author{Robert Lindner}
\affiliation{University of Wisconsin--Madison, Department of Astronomy, 475 N Charter St, Madison, WI 53703, USA}

\author{Yangyang Liu}
\affiliation{University of Wisconsin--Madison, Department of Astronomy, 475 N Charter St, Madison, WI 53703, USA}
\affiliation{current: Department of Pharmacology, College of Veterinary Medicine, Cornell University, Ithaca, 14853, New York, USA}

\author{Naomi McClure-Griffiths} 
\affiliation{Research School of Astronomy \& Astrophysics, The Australian National University, Canberra ACT 2611, Australia}

\author{Tony Wong} 
\affiliation{Department of Astronomy, University of Illinois, 1002 West Green Street, Urbana, IL 61801, USA}

\author{Blair Savage}
\affiliation{University of Wisconsin--Madison, Department of Astronomy, 475 N Charter St, Madison, WI 53703, USA}
\affiliation{Deceased}

\begin{abstract}
The Magellanic Stream, a tidal tail of diffuse gas falling onto the Milky Way, formed by interactions between the Small and Large Magellanic Clouds, is primarily composed of neutral atomic hydrogen (HI). The deficiency of dust and the diffuse nature of the present gas make molecular formation rare and difficult, but if present, could lead to regions potentially suitable for star formation, thereby allowing us to probe conditions of star formation similar to those at high redshifts. We search for \hcop, HCN, HNC, and C$_2$H using the highest sensitivity observations of molecular absorption data from the Atacama Large Millimeter Array (ALMA) to trace these regions, comparing with HI archival data from the Galactic Arecibo L-Band Feed Array (GALFA) HI Survey and the Galactic All Sky Survey (GASS) to compare these environments in the Magellanic Stream to the HI column density threshold for molecular formation in the Milky Way. We also compare the line of sight locations with confirmed locations of stars, molecular hydrogen, and OI detections, though at higher sensitivities than the observations presented here. \\
We find no detections to a 3$\sigma$ significance, despite four sightlines having column densities surpassing the threshold for molecular formation in the diffuse regions of the Milky Way. Here we present our calculations for the upper limits of the column densities of each of these molecular absorption lines, ranging from $3 \times 10^{10}$ to $1 \times 10^{13}$ cm$^{-2}$. The non-detection of \hcop{} suggests that at least one of the following is true: (i) $X_{\hcop{}, \mathrm{MS}}$ is significantly lower than the Milky Way value; (ii) that the widespread diffuse molecular gas observed by \citet{Rybarczyk22b} in the Milky Way's diffuse ISM does not have a direct analog in the MS; (iii) the HI-to-\htwo{} transition occurs in the MS at a higher surface density in the MS than in the LMC or SMC; or (iv) molecular gas exists in the MS, but only in small, dense clumps. 
\end{abstract}

\section{Introduction}
The Magellanic Stream (MS) is a tidal tail of diffuse neutral atomic hydrogen (HI) that formed from interactions of the Small and Large Magellanic Clouds (SMC and LMC) with each other and with the Milky Way. At a metallicity of approximately 10$\%$ of the solar value \citep[][]{Fox13}, the MS is far less chemically enriched than the Milky Way or the LMC (45-50$\%$ solar), though comparable with the SMC  \citep[10--20$\%$ solar,][]{Rolleston02,Fox13}. 
While the part of the MS close to the Clouds has an estimated distance of 50--65 kpc, \citep{Keller06}, distance to the tip of the Stream is observationally not well constrained and ranges from 20 \citep{McClure-Griffiths08, Casetti-Dinescu14} to 100 kpc \citep{Chandra2023}. Other simulations suggest further distances up to 100 kpc \citep{Besla10, Lucchini21} as well. 

To understand the evolution of the interstellar medium (ISM) and the process of star formation, it is important to examine a variety of interstellar environments. A necessary condition for star formation is dense, cold gas, in which self-gravity is able to overcome thermal, turbulent, and magnetic pressures. This gas is usually rich with molecules, particularly molecular hydrogen (H$_2$). Due to the MS's low metallicity and extended, highly diffuse nature, it offers an extremely different laboratory for studying the early stages of molecule formation \citep[and potentially star formation, e.g.,][]{Price-Whelan19} than the chemically enriched, cold, dense, molecular clouds in the plane of the Milky Way.
While the MS is the product of tidal stripping and not a direct analog for sites of primordial star formation, it nevertheless provides a nearby laboratory to study gas in some of the conditions that would have been present at higher redshifts \citep{Murray15, Galametz20}.

The MS has been extensively imaged in HI \citep[e.g.,][]{Putman03}. The HI observations 
of the tip of the MS, the region farthest away from the MCs \citep{Stanimirovic08,Nidever10}, revealed several long filaments, demonstrating that the MS is very extended and with significant small-scale HI structure. 
The wealth of observational data \citep[e.g.,]{Weiner1996, Fox2014, For14, Kim2024} suggests that the MS has a complex multi-phase structure that spans a large range in gas densities and ionization conditions. About 15\% of hydrogen clouds in the sample of \cite{Stanimirovic08} have velocity profiles which suggest two temperature components (one cooler at $<2,000\rm\,K$, and one warmer at $<13,000\rm\,K$). As summarized in \cite{Stanimirovic10}, the existence of the multi-phase medium in the MS, embedded in the ambient hot ($10^6\rm\,K$) Galactic halo gas, is surprising considering Stream's low metallicity, interstellar radiation field, and relevant heating and cooling rates \citep{Wolfire1995,Sternberg2002}. Even more surprisingly, \cite{Matthews2009} detected the first HI absorption lines against radio background sources located behind the MS.  The two detected absorption features suggest temperatures of the absorbing clouds of 70--80\,K, and hydrogen column densities of $2\times 10^{20}\,\rm  cm^{-2}$. \citet{Dempsey20} searched for HI absorption in the far outskirts of the SMC --- where the metallicities are similar to those in the MS \citep[the SMC is thought to be the origin of much of the MS gas;][and references therein]{DonghiaFox2016} --- and detected cold HI with high confidence in two out of five positions. They estimated cold HI properties as having spin temperature 
$\sim 100$ K and column density of $(6-16) \times 10^{20}$ cm$^{-2}$.

Remarkably, molecular hydrogen (\htwo{}) has also been detected within the MS, measured through UV absorption toward the Seyfert galaxy NGC~3783 \citep{Sembach2001,Wakker2006,Richter2013} and toward the quasar Fairall~9 \citep{Richter2001,Wakker2006,Richter18}, in diffuse environments with the HI column density $<10^{20}$ cm$^{-2}$. These observations suggest \htwo{} column densities $N({\rm H_2}) \sim 10^{18}~\persc{}$ and excitation temperatures of $\sim 100~\mathrm{K}$. 
\citet{Lehner02} also detected more diffuse \htwo{} in the direction of the Magellanic Bridge, located between the SMC and the LMC, with $N(\htwo{})=2.7\times10^{15}~\persc{}$.
Together, these observations indicate that molecular hydrogen can form and survive in the diffuse environments that exist in the MS and the Bridge, far from the main bodies of the SMC and LMC.

While the Magellanic Bridge contains young stars \citep{Irwin85}, numerous stellar searches in the MS were unsuccessful \citep[e.g.,][]{Mathewson79}. Only very recently, \cite{Chandra2023} identified 13 stars that are likely located in the MS at a distance of 60--120 kpc. Some of those 13 stars are metal-rich and belong to the recently formed tidal counterpart to the MS, but other stars are  metal-poor and possibly originated from the SMC outskirts during an earlier interaction between the Clouds. \cite{Zaritsky20} also detected 15 stars of SMC metallicity with velocities of the MS likely formed by a tidal interaction of the MS with other compact clouds in the outer edges of the MW.
Following the star formation models, we would expect that molecular gas can form and survive in the MS if some of star formation has indeed occurred there.

Previous direct detections of \htwo{} in the MS have been valuable for characterizing molecule formation in this extreme environment, but these measurements have been extremely rare because molecular hydrogen lacks a permanent electric dipole moment, making it extremely difficult to detect in diffuse environments such as the MS. 
Instead, observations from other molecular species --- which are much easier to detect in emission and absorption --- are most often used to trace the molecular content of the gas in these environments.
While carbon monoxide (CO)---the most common tracer of molecular gas in galaxies---has been detected in emission in several positions in the Magellanic Bridge \citep{Muller_2003,Mizuno2006}, no detections of CO emission exist for the MS. Most metallicity estimates suggest that the MS was formed out of the SMC gas, however metallicity variations along the MS suggest some mixing of the SMC/LMC gas \citep{Fox13,Richter2013}. Several studies  \citep[e.g.,][]{Richter2013} suggest that the MS likely has a clumpy but widespread diffuse molecular phase that is enabled by efficient H$_2$ self-shielding and the absence of local UV sources \citep{Fox2005}. 

In the Milky Way, observations of the molecule \hcop{} in absorption are commonly used to trace diffuse molecular gas. \hcop{} is readily detected in absorption in directions with $A_V\gtrsim0.25~\mathrm{mag}$ \citep[e.g.,][]{LucasLiszt1996, Rybarczyk22b}, evidently forming under similar conditions to the HI-to-\htwo{} transition. 
Moreover, \hcop{} absorption is regularly detected  in directions where CO emission is undetected \citep{LucasLiszt1996,LisztLucas2000,Rybarczyk22b}. 
This suggests that a search for \hcop{} in absorption is among the best approaches for detecting molecular gas in the diffuse environment of MS.
Indeed, \citet{Murray15} reported a tentative detection of \hcop{} absorption in the Magellanic system outside of the LMC and SMC, in the low density environment of the Magellanic Bridge. 

In this study, we have obtained the most sensitive observations to date of molecular absorption from key molecular species, including \hcop{}, in the MS using the the Atacama Large Millimeter/submillimeter Array (ALMA). 
In Section \ref{Observations}, we discuss the observations for each of the four molecular lines, HCO$^+$, HCN, HNC, and C$_2$H at 10 pointings in the Magellanic Stream.
In Section \ref{Upper Limits}, we present the upper limits on the optical depths ($\tau$) and the column densities of the four molecular species for each line of sight. 
In Section \ref{Spatial_dist}, we examine the position of each line of sight relative to the SMC and LMC to contextualize the column density limits presented in Section \ref{Upper Limits}. We also search for possible correlations between the location conditions and the upper limits of the optical depths and column densities of different molecular species.
In Section \ref{sec:fmol}, we discuss what our observations imply for the
molecular fraction in the Stream.

\section{Observations}
\label{Observations}
\subsection{Molecular absorption  measurements from ALMA}
We searched for absorption by the $J = 1-0$ transitions of HCN (including three hyperfine lines at $88.6304$--$88.6339~\mathrm{GHz}$), HCO$^+$ ($89.1890~\mathrm{GHz}$), and HNC ($90.6336~\mathrm{GHz}$), and the $N = 1-0$ transition of C$_2$H (including two hyperfine lines at $87.3169~\mathrm{GHz}$ and $87.3286~\mathrm{GHz}$) using ALMA Band 3 observations toward 14 background radio continuum sources (project code 2013.1.00700.S, ALMA Cycle 2). Observations were obtained between May 2015 and September 2015. On-source integration times ranged from 5--10 minutes.

Three background sources---NGC3783, NGC7714 and NGC7469---were chosen because they had complementary UV observations. Unfortunately, all three sources were excluded from our analysis because their continuum flux densities were too low to be able to detect absorption (0.003 Jy, 0.002 Jy, and 0.003 Jy, respectively).
We chose the remaining ten background sources based on their positions behind the MS and their high continuum flux densities. Table \ref{tab_coordinates} lists the positions and 89~GHz flux densities of all ten sources included in our analysis. 
The velocity resolutions of the final absorption-line spectra ranged from 0.303 km s$^{-1}$ to 0.315 km s$^{-1}$.

\begin{table}[t!]
	\begin{center}
	\caption{Summary of observations. For each source (column 1), we list the celestial coordinates (columns 2 and 3), the Magellanic Stream coordinates (columns 4 and 5), the flux density at 89 GHz (column 6), the optical depth sensitivity for HI absorption using ATCA (Column 7), and the optical depth sensitivity in the final spectra achieved for each of the four molecular transitions observed with ALMA (columns 8--11).}
	\label{tab_coordinates}
	\begin{tabular}{l | c | c | c | c | c | c | c | c | c }
\hline\hline
Source & $\alpha (^\circ)$ & $\delta (^\circ)$ &$l_{\mathrm{MS}}$ & $b_{\mathrm{MS}}$ & $F_{89}~(\mathrm{Jy})$ &  $\sigma_{\tau,\mathrm{HCO^+}}$ & $\sigma_{\tau,\mathrm{HCN}}$ & $\sigma_{\tau,\mathrm{HNC}}$ & $\sigma_{\tau,\mathrm{C_2H}}$ \\
\hline
2331+073   & 353.553 & 7.608 & -96.29 & 1.62 & 0.4683 & 0.023 & 0.021 & 0.023 & 0.020 \\
J2230+114  & 338.152 & 11.731 & -105.22 & -11.37 & 3.014 & 0.004 & 0.004 & 0.004 & 0.004 \\
J0635-7516 & 98.944 & -75.271 & 4.87 & -3.51 & 1.264 & 0.008 & 0.008 & 0.009 & 0.008 \\
2355-534   & 359.472 & -53.187 & -36.28 & -11.13 & 0.9927 & 0.010 & 0.010 & 0.010 & 0.010 \\
J1058-8003 & 164.680 & -80.065 & 11.36 & -15.87 & 1.4230 & 0.017 & 0.010 & 0.016 & 0.014 \\
J0454-8101 & 72.523 & -81.017 & -0.67 & -8.98 & 2.0043 & 0.009 & 0.008 & 0.008 & 0.007 \\
2345-167   & 357.011 & -16.520 & -72.27 & -2.50 & 3.0402 & 0.005 & 0.005 & 0.005 & 0.004 \\
J0049-4457 & 12.319 & -44.953 & -41.12 & -0.42 & 0.02834 & 0.034 & 0.040 & 0.038 & 0.039 \\
J0253-5441 & 43.372 & -54.698 & -21.40 & 9.58 & 1.1720 & 0.012 & 0.010 & 0.020 & 0.011 \\
2326-477   & 352.324 & -47.505 & -42.99 & -14.32 & 0.71428 & 0.019 & 0.018 & 0.017 & 0.015 \\

	\end{tabular}
	\end{center}
\end{table}

The calibration and imaging for these sources were straightforward, and the ALMA pipeline products were essentially science-ready. To obtain the most accurate estimates of the continuum flux densities, though, we reran the ALMA pipeline without continuum subtraction for all of our observations. 
Toward each background source, we then extracted the absorption spectrum,
\begin{equation} \label{eq:emtau}
    e^{-\tau(v)} = \frac{I(v)}{I_0(v)},
\end{equation}
where $\tau(v)$ is the optical depth, $I(v)$ is the observed intensity, and $I_0(v)$ is the continuum emission of the background source. All three quantities are functions of frequency (here we convert to radial velocity, $v$).
For each molecular line, we selected the pixel with the brightest continuum flux density and extracted the spectrum ($I(v)$ in Equation \ref{eq:emtau}) from the spectral cube at that pixel. To model the continuum ($I_0(v)$ in Equation \ref{eq:emtau}), we fit a constant value using the spectral channels where HI is not seen in emission, and therefore no molecular absorption is expected \citep[e.g.,][]{Rybarczyk22b}. We chose to fit a constant because the continuum was nearly flat across each of the $58.50~\mathrm{MHz}$ spectral windows. 

\begin{figure}[h!]
	\centering
        \includegraphics[width=0.9\textwidth]{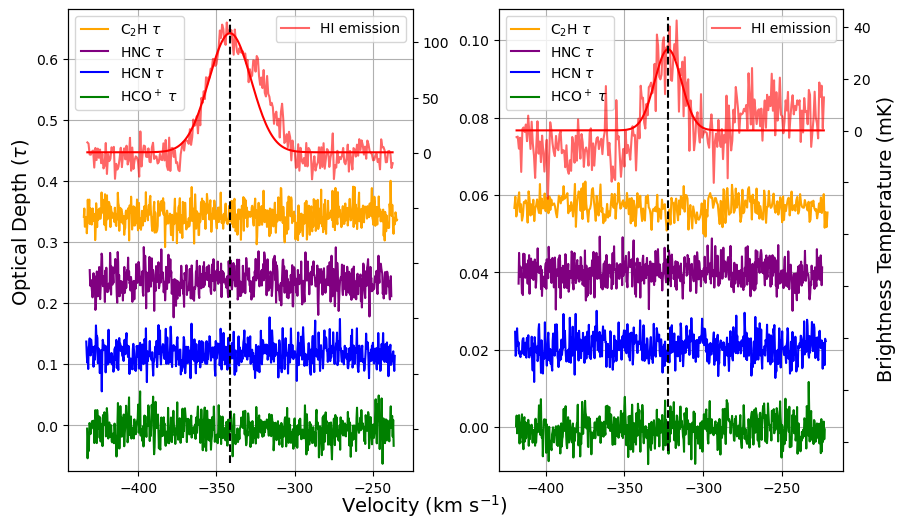}\\
	\caption{The optical depth spectra for the sources 2331+073 (left) and J2230+114 (right) with the corresponding HI brightness temperature spectra from the GALFA-HI survey.}
	\label{fig_galfa_spectra}
\end{figure}

\begin{figure}[h!]
	\centering
        \includegraphics[width=\textwidth]{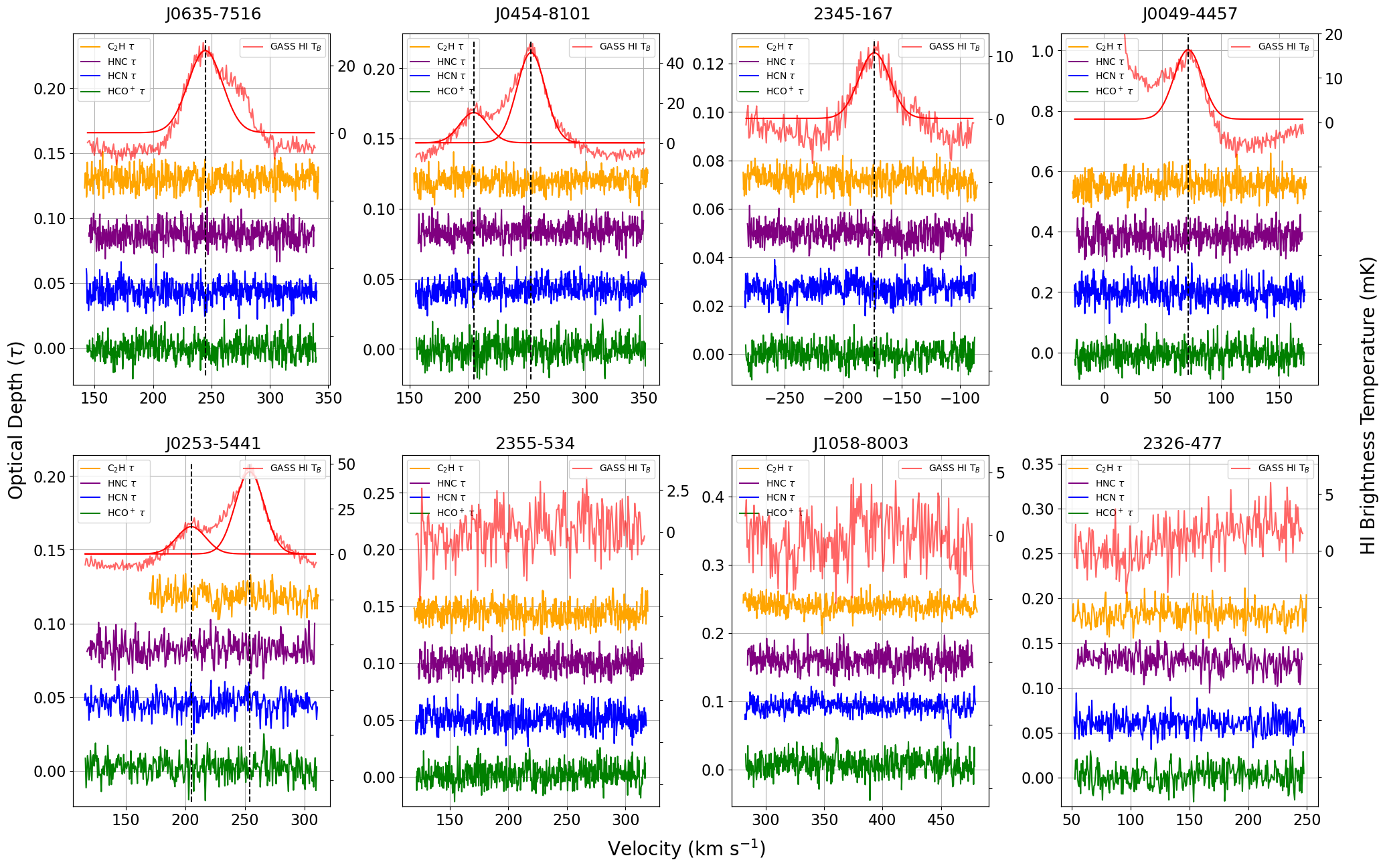}\\
	\caption{The optical depth spectra for the sources J0635-7516, J0454-8101, 2345-167, J0049-4457, J0253-5441, 2355-534, J1058-8003, and 2326-477 with their corresponding HI brightness temperature spectra from GASS.}
	\label{fig_gass_spectra}
\end{figure}

In Figures \ref{fig_galfa_spectra} and \ref{fig_gass_spectra}, we show the optical depth spectra, $\tau(v)$, measured toward each background source. From bottom to top, we plot the optical depth spectra of HCO$^+$ (green), HCN (blue), HNC (purple), C$_2$H (yellow).

\subsection{Complementary HI measurements}
In Figures \ref{fig_galfa_spectra} and \ref{fig_gass_spectra}, we show the HI emission spectrum (red, top) alongside the optical depth spectra measured in the direction of each background source. Figure \ref{fig_galfa_spectra} uses HI emission spectra measured by the Galactic Arecibo L-band Feed Array HI (GALFA-HI) Survey \citep{Peek11,Peek18}.
The HI spectra in Figure \ref{fig_gass_spectra} come from the Galactic All Sky Survey \citep[GASS,][]{McClure_Griffiths09} and the HI4PI survey \citep{HI4PI}.
The 1$\sigma_{rms}$ noise level is 43 mK, 55 mK, and 60 mK -- 140 mK for the GASS, HI4PI, and GALFA-HI surveys, respectively \citep{HI4PI, McClure_Griffiths09, Peek18}, corresponding to 1$\sigma_{rms}$ $N(\mathrm{HI})$ sensitivities $\sim 2.3 \times 10^{18} \mathrm{cm}^{-2}$, $2.5 \times 10^{18} \mathrm{cm}^{-2}$, and $6.5 \times 10^{18} \mathrm{cm}^{-2}$, assuming a FWHM of 20 km s$^{-1}$, 30 km s$^{-1}$, and 20 km s$^{-1}$, respectively.
For each sightline, we fit a Gaussian function to the main emission feature.
The centroid velocity of the fit is shown
with the black dashed vertical lines in Figures \ref{fig_galfa_spectra} and \ref{fig_gass_spectra} to emphasize at which velocities molecular absorption lines would be expected if present. 

Molecular gas forms from relatively cold and dense HI; if cold molecular gas were present in any of these directions, we would also expect cold HI to be present and potentially detectable in absorption \citep[e.g.,][]{Stanimirovic14, Nguyen19, Rybarczyk22b, Park23, Hafner23}. Moreover, HI absorption observations would probe a similar physical scale to our pencil-beam molecular absorption line spectra, whereas HI emission observations probe a much larger physical scale. \cite{Murray15} observed several of the same sources observed in this study in several wavelengths including HI absorption. These yielded all non-detections for the observed sources. Because we detect no HI gas in absorption, our assumption made in Section \ref{Upper Limits} is valid and we derive all HI column densities and velocities from the HI emission spectra. \\

\section{Limits on the molecular column densities in the MS}
\label{Upper Limits}

We do not detect \hcop{}, C$_2$H, HCN, or HNC absorption in the direction of any of the 10 background sources listed in Table \ref{tab_coordinates} at a level of $3\sigma$. In Table \ref{tab_coordinates}, we list the $1\sigma$ noise in $\tau(v)$ for velocity channels of width $\sim 0.3 ~ \mathrm{km\ s}^{-1}$ for each transition and each line of sight. Our sensitivity is comparable to that achieved by \citet{Rybarczyk22a}, who detected \hcop{} absorption in 10 diffuse sightlines through the Milky Way. 
If we assume a conversion factor $X_{\hcop{}}\sim\mathrm{few}\times10^{-9}$ as suggested by
\citet{MillarHerbst1990} for both the SMC and LMC, our upper limit on the H$_2$ column density is $\sim10^{19}$ cm$^{-2}$ (although see discussion of \hcop{} abundance in Section \ref{sec:fmol}).

\begin{table}[h]
    \centering
    \caption{Conversion factors from integrated optical depth to column density ($C(T_\mathrm{ex})$ in Equation \ref{eq:tau_to_N}) for the transitions observed in this work, assuming $T_{\mathrm{ex}}=2.725~\mathrm{K}$ \citep[e.g.,][]{Godard2010,Luo2020}.}
    \begin{tabular}{c | c}
    \hline \hline
        Species &  $C(T_{ex}) = N/\int \tau dv~[\mathrm{cm^{-2}}/\mathrm{km~s^{-1}}]$\\
        \hline
        C$_2$H & $4.34\times10^{13}$ \\
        HCN & $1.91\times10^{12}$ \\
        HCO$^+$ & $1.11\times10^{12}$ \\
        HNC & $1.78\times10^{12}$ 
    \end{tabular}
    \label{tab:tau_to_N}
\end{table}

To estimate upper limits to the column densities of HCO$^+$, HCN, HNC, and C$_2$H we make several assumptions about the nature of a molecular line detection in this environment. The column density, $N$, of a molecular species is related to the optical depth, $\tau(v)$, by
\begin{equation} \label{eq:tau_to_N}
    N = C(T_{\mathrm{ex}}) \int \tau(v) dv,
\end{equation}
where $C$ depends on the the excitation temperature, $T_{\mathrm{ex}}$, and the physical properties of the observed transition \citep[e.g.,][]{DraineTextbook}.
We assume an excitation temperature equal to the temperature of the CMB, $2.725~\mathrm{K}$, for all lines, as is typically assumed in diffuse environments \citep[e.g.,][]{LucasLiszt1996,LucasLiszt2000_CCH,LisztLucas2001_HCN_HNC} and has been confirmed in some diffuse directions for \hcop{} \citep[][]{Godard2010,Luo2020}. Values for $C(2.725~\mathrm{K})$ are listed in Table \ref{tab:tau_to_N}. 
We also assume that any molecular absorption lines are Gaussian, with peak optical depth $\tau_0$ and full width at half maximum (FWHM) $\Delta v_0$.
Then, the column density is $N=C(T_{\mathrm{ex}}) \times 1.064\times \tau_0 \times \Delta v_0$. 
We assume a minimum FWHM of 2 $\kms{}$ \citep[e.g.,][]{LucasLiszt1996,Godard2010,Rybarczyk22a}. Since we find only non-detections, in Table \ref{tab_cd_upper_limits} we report upper limits to the molecular column densities using $\tau_0=3\sigma_\tau$.

We also compare these molecular column density upper limits with the HI column densities and surface densities (Table \ref{tab_cd_upper_limits}). We find the column densities of the HI using the integrated intensities averaged over 20'' when using the GALFA-HI data and 44'' when using the GASS data to account for diffuse larger structures in that region. The sizes of the areas differed between the two HI surveys because the angular resolution was far improved in the GALFA-HI survey compared to the GASS survey, which allowed the HI emission in the Magellanic Stream to be resolved on smaller angular scales. However because the intensities were averaged, we find minimal differences in the outcomes. These values are then multiplied by a scale factor of $1.823 \times 10^{18}~\persc/(\mathrm{K~km~s^{-1}})$ \citep{Murray18}. This assumes an  optically thin environment, so these represent lower limits of the HI column densities. We then use a scale factor of $0.8 \times 10^{-20}$ to convert between the HI column densities, $N(\mathrm{HI})$, and the HI surface densities, $\Sigma(\mathrm{HI}) (\mathrm{M}_\odot \mathrm{pc}^{-2})$ \citep{Roman-Duval14}.  

\begin{table}[h!]
	\begin{center}
	\caption{HI column density and surface density and upper limits ($3\sigma$) of the column densities of HCO$^{+}$, HCN, HNC, and C$_2$H for each line of sight in our sample.}
	\label{tab_cd_upper_limits}
	\begin{tabular}{c | c | c | c | c | c | c} 
\hline\hline
Position & $N(\mathrm{HI})$ & $\Sigma (\mathrm{HI})$ & $N(\mathrm{HCO^+})$  & $N(\mathrm{HCN})$ & $N(\mathrm{HNC})$ & $N(\mathrm{C_2H})$ \\
		  & $10^{17} ~\mathrm{cm^{-2}}$ & $M_{\odot}~\mathrm{pc}^{-2}$ & $10^{11}~\mathrm{cm^{-2}}$ & $10^{11}~\mathrm{cm^{-2}}$ & $10^{11}~\mathrm{cm^{-2}}$ & $10^{12}~\mathrm{cm^{-2}}$ \\
\hline
2331+073    & $180.5$ & $ 0.14$ & $<1.636$ & $<2.504$ & $<2.490$ & $<5.428$   \\
J2230+114   & $45.5$ & $0.036$ & $<0.290$ & $<0.442$ & $<0.411$ & $<1.029$ \\
J0635-7516  & $140.0$ & $0.11$ & $<0.587$ & $<1.006$ & $<1.051$ & $<2.387$  \\
2355-534    &  $4.1$ & $0.0033$ & $<0.725$ & $<1.185$ & $<1.155$ & $<2.725$ \\
J1058-8003  & $<0.3$ & $<0.0003$ & $<1.211$ & $<1.242$ & $<1.778$ & $<3.628$ \\
J0454-8101  & $190.6$ & $0.15$ & $<0.625$ & $<0.983$ & $<0.863$ & $<1.874$  \\
2345-167    & $<38.2$ & $<0.031$ & $<0.327$ & $<0.565$ & $<0.522$ & $<1.367$  \\
J0049-4457  & $30.7$ & $0.025$ & $<2.392$ & $<4.999$ & $<4.330$ & $<10.774$  \\
J0253-5441  & $306.6$ & $0.25$ & $<0.843$ & $<1.374$ & $<1.461$ & $<2.969$ \\
2326-477    & $<30.6$ & $<0.025$ & $<1.265$ & $<2.035$ & $<2.040$ & $<4.792$ \\
\hline
	\end{tabular}
	\end{center}
\end{table}

We compare these results with those of \cite{Murray15}, who found a tentative detection of HCO$^+$ toward J0454-8101 with a peak $\tau$ of $0.10 \pm 0.02$ with $\sigma_{\tau} =  0.026$ 
per 0.8 km s$^{-1}$ velocity channels using the Australia Telescope Compact Array (ATCA). For HCO$^+$ our value at J0454-8101 was $\sigma_{\tau} = 0.009$, which would make our 3$\sigma$ upper limit equal to $\tau = 0.027$. We are unable to reproduce the results of \cite{Murray15} here. 
The tentative detections from \cite{Murray15} could be a result of radio frequency interference (RFI) contamination. The location of ALMA should be less susceptible to RFI contamination than ATCA, and our observations have both a higher sensitivity and an increased bandpass stability.

\begin{figure}[h]
	\centering
        \includegraphics[width=0.8\textwidth]{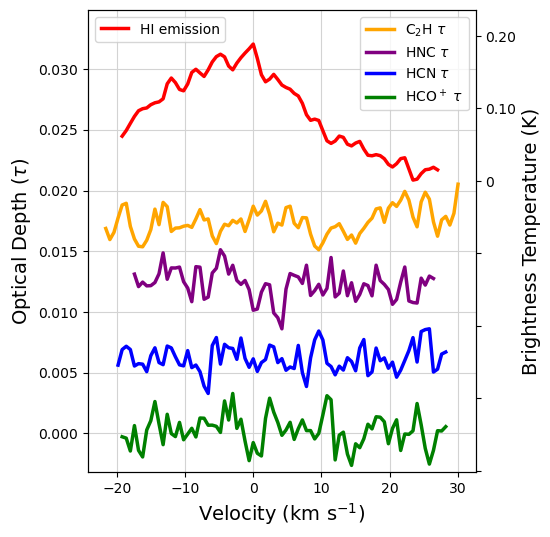}\\
	\caption{The smoothed and stacked spectra for each of the molecular species and the HI emission spectrum from GASS and GALFA}
	\label{fig_stacked_spectra}
\end{figure}

Finally, in an effort to improve the signal-to-noise ratio in our molecular absorption spectra, we stack all 10 spectra for each molecular absorption line to create a weighted mean spectrum (Figure \ref{fig_stacked_spectra}). The spectra are weighted by the inverse square of the standard deviation of individual spectra. When stacking, all of the composite spectra are shifted in velocity so that $0~\kms{}$ in each stacked spectrum aligns with the peak in the HI emission spectrum\footnote{Ideally, we would align the spectra along the velocity of peak HI absorption, since we expect molecular absorption to be associated with cold HI. Because we do not detect HI absorption in these directions, though, we use the HI emission instead. Figures \ref{fig_galfa_spectra} and \ref{fig_gass_spectra} show that the HI emission spectra are quite simple in these directions, with only one or two components.}. The optical depth sensitivities for each of the stacked molecular absorption spectra are listed in Table \ref{tab:stacked_spectra}.
Stacking improves the signal to noise ratio by a factor of 1.5 -- 10 compared to the individual spectra shown in Figures \ref{fig_galfa_spectra} and \ref{fig_gass_spectra}.
Even with the reduced noise in the stacked spectra, though, we still do not detect molecular absorption at a level 3$\sigma$ in any of the four lines. 
Upper limits to the molecular column densities in the stacked spectra are listed in Table \ref{tab:stacked_spectra}.

\begin{table}[h]
	\begin{center}
	\caption{Upper limits ($3\sigma$) of the column densities for the stacked spectra of HCO$^{+}$, HCN, HNC, and C$_2$H.}
	\label{tab:stacked_spectra}
	\begin{tabular}{c | c | c} 
\hline\hline
Molecular Species & Column Density $(\mathrm{cm^{-2}})$ & $\sigma_\tau$  \\
\hline
$\mathrm{HCO^+}$ & $<9.257\times 10^{9}$ & $0.0013$ \\
HCN   & $<1.644\times 10^{10}$ & $0.0012$ \\
HNC   &  $<1.226\times 10^{10}$ & $0.0010$ \\
$\mathrm{C_2H}$  & $<3.478 \times 10^{11}$ & $0.0012$ \\
HI    & $8.100 \times 10^{18}$ & - \\
\hline
	\end{tabular}
	\end{center}
\end{table}

\section{Discussion}
\label{Mapping}

\subsection{Spatial distribution of molecular upper limits across the Magellanic Stream}
\label{Spatial_dist}

\begin{figure}[h!]
	\centering
        \includegraphics[width=0.9\textwidth]{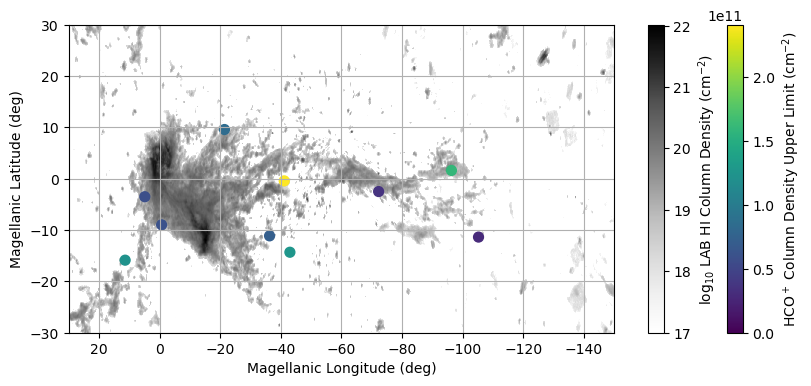}
	\caption{The Upper Limits of the Column Densities of HCO$^+$ plotted against the HI column density map of the high velocity components ($\pm 100$ km s$^{-1}$) of the Magellanic Stream using the Magellanic Stream Coordinate System. The image is from the LAB High Velocity Cloud Sky Survey and has angular resolution of $36'$ \citep{Westmeier18}.}
	\label{fig_stream}
\end{figure}

In Figure \ref{fig_stream}, we show the locations of our background sources across the Magellanic Stream, using a color bar to indicate the upper limits of the column density of each molecular line. 
The background of Figure \ref{fig_stream} shows the column density of the high velocity ($|v_{LSR}| < 450$ km s$^{-1}$, $v_{dev}$ = 70 km s$^{-1}$) HI emission in the Magellanic Stream and Leading Arm from the LAB (Leiden/Argentine/Bonn) survey \citep{Westmeier18}, calculated using the HI4PI all sky survey \citep{HI4PI}. The HI column density is shown using the Magellanic Stream Coordinate System \citep{Nidever08}. 

We transformed the coordinates of each of the background sources to the Magellanic Stream Coordinate system to emphasize the Magellanic Stream and Leading Arm in the HI4PI data. This was accomplished following the equations first introduced by \cite{Wakker01} and \cite{Nidever08}, 
\begin{equation}
    \tan(\lambda-\lambda_0) = \frac{\sin(b) \sin(\epsilon) + \cos(b) \cos(\epsilon)\sin(l-l_0)}{\cos(b)\cos(l-l_0)}
    \label{eq_MS_long}
\end{equation}

\begin{equation}
   \sin(\beta) = \sin(b)\cos(\epsilon) - \cos(b)\sin(\epsilon)\sin(l-l_0),
    \label{eq_MS_lat}
\end{equation}
where $l$ and $b$ are the Galactic latitude and longitude values, which return $\lambda$ and $\beta$, the Magellanic longitude and latitude values. 
In these equations, $l_0 = 278.5^\circ$, $\lambda_0 = 32.861^\circ$, and $\epsilon = 97.5^\circ$ \citep{Nidever08}.


As shown in Figure \ref{fig_stream}, we probe a range of HI column density environments. Two sources probe gas in the outskirts of the Magellanic Clouds with higher HI column densities ($2 \times 10^{19}$ cm$^{-2}$) and less filamentary and dispersed gas structures. Three sources probe the Leading Arm (gas on the opposite side of the Stream). At one of these sources (J1058-8003), we do not detect HI and therefore provide an upper limit to the column density as seen in Table \ref{tab_cd_upper_limits}. The remaining seven sources are in the direction of the Stream with varied column densities ranging from $4 \times 10^{17}$ -- $3 \times 10^{19}$ cm$^{-2}$, though at two of these sources (2345-167 and 2326-477) we do not detect HI emission. Upper limits for the column densities can again be found in Table \ref{tab_cd_upper_limits}. 
Our sensitivity is roughly uniform across our sample. For most sources, we find $N(\hcop{})\lesssim\mathrm{few}\times10^{10}~\persc{}$, both around the MCs and in the leading arm of the MS. 

Several previous studies have detected H$_2$ using UV spectroscopy in the direction of the MS \citep{Sembach2001,Richter2001,Wakker2006,Richter2013,Richter18} and Magellanic Bridge \citep{Lehner02}. \citet{Sembach2001}, \citet{Wakker2006}, and \citet{Richter18} detected \htwo{} and other species toward the MS Leading Arm, while \citet{Richter2001}, \citet{Wakker2006}, and \citet{Richter2013} detected \htwo{} and other species toward the MS. In both directions, best estimates for the \htwo{} column densities are $N(\htwo{})\sim10^{18}~\persc{}$, with $f_{\mathrm{mol}}=2n(\htwo{})/n_{\mathrm{H}}\approx0.01$--$0.04$. 
\htwo{} has also been detected in the Magellanic Bridge \citep{Lehner02}, but with relatively lower column density, $N(\htwo)=2.7\times10^{15}~\persc$, and lower molecular fraction, $f_{\mathrm{mol}}\approx10^{-5}$--$10^{-4}$.
In Section \ref{sec:fmol} below, we discuss our non-detections in the context of these previous results.

\begin{figure}[h!]
    \centering
    \includegraphics[width=0.9\textwidth]{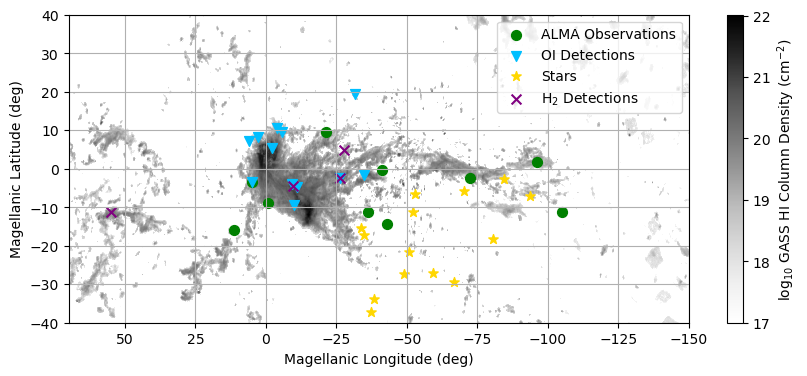}
    \caption{A spatial comparison between detections of O I (blue triangles), stars (yellow stars), \htwo (purple xs), and our observations (green circles) in the Magellanic Stream and Leading Arm, plotted in the background using data from the LAB High Velocity Cloud Sky Survey.}
    \label{fig:stars}
\end{figure}

Recently, several studies claimed the detection of stars along and around the MS. \citet{Chandra2023} detected 13 stars, of ages 4-10 Gyr, trailing the past orbits of the Clouds. Based on kinematics and metallicity these are high-confidence members of the MS. These stars, which we show in Figure~\ref{fig:stars}, have two populations: about half of the stars form a higher-metallicity population that tracks the HI well, and the rest of the stars form a diffuse, lower-metallicity population that is offset from the HI by about 20 degrees. The higher-metallicity stars were most likely formed in the LMC disk and were stripped at the same time as the HI gas. However, if formed in-situ, as was suggested for 1-4 Gyr-old stars discovered in the Milky Way Halo \citep{Price-Whelan19}, thought to have formed during the last interaction of the Leading Arm, these stars in the Magellanic Stream would need cold and dense molecular gas to form. The origin of the lower-metallicity stars is harder to constrain, with one idea being that those stars were thrown out of the SMC outskirts during past interactions between the Clouds. If some of recently detected stars were formed in-situ, \cite{Price-Whelan19} proposes this to have occurred with tidally stripped gas interacting with Milky Way halo gas during recent low-mass mergers. During these interactions, the turbulence of the infalling gas could have created cold and dense regions where some star formation may have occurred \citep[see][and references therein]{MacLowKlessen2004}; molecular gas would also be formed in this process. The existence of stars in the Magellanic Stream therefore suggests the possibility of the presence of molecular gas, which we constrain here.

Using the Cosmic Origins Spectrograph (COS) on the Hubble Space Telescope, recent studies observed OI in the Magellanic Corona, surrounding the Magellanic Clouds and into the Magellanic Stream \citep{Krishnarao22}. These 12 detections were found to be consistent with tidally stripped gas from the Magallanic Stream rather than originating in the more higher temperature ($10^6$ K) and highly ionized Magellanic Corona \citep{Krishnarao22}. The other ionized species detected with COS indicated a higher correlation with LMC latitudes, suggesting origins from impacts with the LMC, while the OI correlated stronger with the Magellanic Stream \citep{Krishnarao22}. 

In Figure \ref{fig:stars}, we show a spatial comparison of the sightlines observed in this work to directions with previously detected \htwo{}, OI, and stars. From this figure, we see that while some of the observed targets probe similar environments, many do not. The majority of the OI detections \citep{Krishnarao22} probe regions closer to the Magellanic Clouds and the Magellanic Corona, which shows higher HI column densities and in several cases correspond spatially with detections of \htwo \citep{Richter2001, Sembach2001, Lehner02}. Most of the detected stars \citep{Chandra2023} do not trace similar regions of the MS compared to our observations, or any of the detections of OI or \htwo. 

\subsection{Constraints on the Molecular Fraction} 
\label{sec:fmol}
In the diffuse ISM of the Milky Way, the abundance of \hcop{} with respect to \htwo{} is roughly constant, with only a limited scatter, $N(\hcop{})/N(\htwo{})=3\times10^{-9}\pm0.2~\mathrm{dex}$ \citep{Gerin2019,LisztGerin2023}. 
If \hcop{} is a similarly good tracer of \htwo{} in the MS, then we can constrain $N(\htwo{})$ and set an upper limit on the molecular fraction of the gas in our MS sightlines,
\begin{equation} \label{eq:fmol}
    f_{\mathrm{mol}} = \frac{2N(\hcop)/X_{\hcop{}, \mathrm{MS}}}{N(\mathrm{HI})+2N(\hcop)/X_{\hcop{}, \mathrm{MS}}} = \Big(1+\frac{N(\mathrm{HI})X_{\hcop{}, \mathrm{MS}}}{2N(\hcop)}\Big)^{-1},
\end{equation}
where $X_{\hcop{}, \mathrm{MS}}=N(\hcop{})/N(\htwo{})$ in the MS.
We note that our estimates for $N(\hcop{})$ come from pencil-beam absorption spectra, whereas our estimates for $N(\mathrm{HI})$ come from emission spectra with much larger beam sizes. Ideally, we would compare our molecular absorption observations to HI absorption observations in order to probe similar physical scales. However, because we do not detect HI in absorption, we use the emission observations to  estimate the total column density of the atomic hydrogen. Equation \ref{eq:fmol} implicitly assumes that $N(\mathrm{HI})$ and $N(\htwo{})$ are probing the same environment, so our non-uniform approach allows us only to make an approximate measurement of the relationship between $f_{\mathrm{mol}}$ and $X_{\hcop{}, \mathrm{MS}}$.

\begin{figure}[h!]
    \centering
    \includegraphics[width=0.9\textwidth]{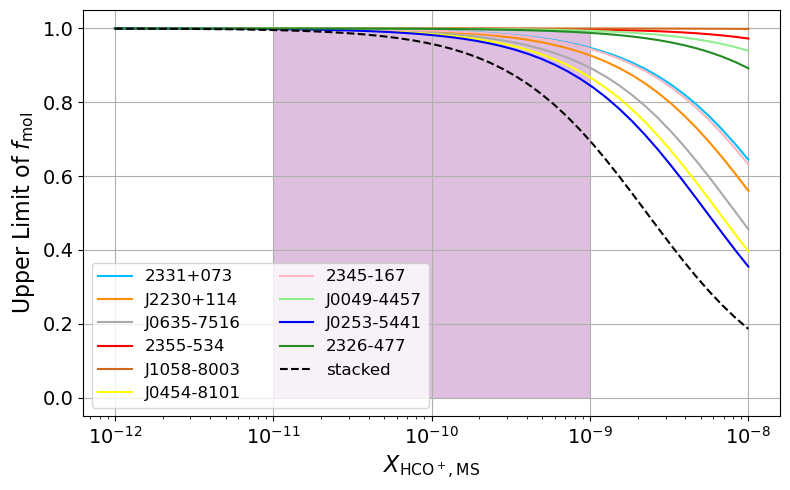}
    \caption{The upper limit on $f_{\mathrm{mol}}$ as a function of $X_{\hcop{},\mathrm{MS}}$ (Equation \ref{eq:fmol}) for each source.}
    \label{fig:fmol}
\end{figure}

In Figure \ref{fig:fmol}, we show what our measurements of $N(\mathrm{HI})$ and $N(\hcop{})$ imply for $f_\mathrm{mol}$ for a range of possible $X_{\hcop{}, \mathrm{MS}}$. To properly interpret our results, we need some reasonable estimate of $X_{\hcop{}, \mathrm{MS}}$.
\citet{MillarHerbst1990} suggest an abundance of $X_{\hcop{}}\sim\mathrm{few}\times10^{-9}$ for both the SMC and LMC, similar to that observed in the Milky Way's diffuse ISM, although their analysis focuses on dark clouds which may be much denser than the diffuse environments of the Stream and therefore have a different $X_{\hcop{}}$ \citep[e.g.,][]{Panessa2023}. We further note that \citet{MillarHerbst1990} estimates are probably inflated due to the low estimate of the H$_3^+$ dissociative recombination rate \citep[see summary by][and references therein]{Larsson2008}. The MS has a lower density than dark clouds and a lower metallicity than the LMC. The relationships of $X_{\hcop{}}$ with density and with metallicity are non-linear, so it is difficult to estimate $X_{\hcop{}, \mathrm{MS}}$ from SMC and LMC analogs. If we use a PDR model \citep{Gong2017} with conservative estimates for the cosmic ray ionization rate, temperature, density, metallicity, and interstellar radiation field of gas in the MS \citep[motivated by direct measurements or comparisons to the SMC/LMC, e.g.,][]{Fox13,Barger2017,Dempsey20,Kosenko2023}, we find a range of $X_{\hcop{}}$ values between $\mathcal{O}(10^{-12})$ and $\mathcal{O}(10^{-10})$. However, the fixed abundance of \hcop{} in the Milky Way \citep{Gerin2019,LisztGerin2023} is an empirical result whose origin remains poorly understood. This result is not predicted by equilibrium chemical models, and PDR models generally underpredict the abundance of \hcop{} in diffuse environments \citep[e.g.,][]{Godard2010}. Given these considerations, we take the range of $10^{-11} \lesssim X_{\hcop{}, \mathrm{MS}} \lesssim 10^{-9}$ to be plausible (this region is shaded in Figure \ref{fig:fmol}), but we consider an even wider range of possible $X_{\hcop{}, \mathrm{MS}}$ values in Figure \ref{fig:fmol}.
Because $N(\hcop{})$ is an upper limit for each line of sight, the curves in Figure \ref{fig:fmol} indicate upper limits to $f_{\mathrm{mol}}$ as a function of $X_{\hcop{},\mathrm{MS}}$.

If $X_{\hcop{}, \mathrm{MS}}\lesssim10^{-10}$, our observations do not give us tight constraints on the molecular fraction beyond indicating that the gas is not fully molecular. If $10^{-10} \lesssim X_{\hcop{}, \mathrm{MS}} \lesssim 10^{-9}$, we can place only slightly better constraints on the sightlines with the highest HI column densities --- in this case, the gas toward J0253-5441, J0454-8101, and J0635-7516 must be $<90\%$ molecular, while the gas toward J2230+114 and J1058-8003 must be $<95\%$ molecular. 
Despite our excellent \hcop{} optical depth sensitivity, the low column densities of the gas in the direction of our background sources mean that, for any plausible $X_{\hcop{}, \mathrm{MS}}$, we can only place weak limits on the fraction of molecular gas toward our 10 lines of sight. 
If we consider the stacked spectrum, where we achieve an \hcop{} optical depth sensitivity $\sigma_\tau\sim0.001$ (Table \ref{tab:stacked_spectra}), we can place somewhat more stringent limits on the molecular fraction: for $X_{\hcop{}, \mathrm{MS}}\lesssim10^{-10}$, we can still only say that the MS gas is less than $\sim95\%$ molecular, but for $10^{-10} \lesssim X_{\hcop{}, \mathrm{MS}} \lesssim 10^{-9}$, we find upper limits on the molecular fraction of $\sim70$--$90\%$.

Previously, direct measurements of \htwo{} using UV absorption  have placed tighter constraints on the molecular fraction in different parts of the Magellanic System. 
In the direction of the MS, toward the quasar Fairall~9, observations of \htwo{} and other neutral species suggest $f_{\mathrm{mol}}=0.01$ \citep{Richter2001,Wakker2006,Richter2013}. Meanwhile, in the direction of the Leading Arm of the MS, toward the Seyfert galaxy NGC~3783, observations suggest $f_{\mathrm{mol}}=0.04$ \citep{Sembach2001,Wakker2006,Richter2013}. 
In the Magellanic Bridge, observations suggest a lower molecular fraction, $f_{\mathrm{mol}}=10^{-5}$--$10^{-4}$ \citep{Lehner02}.

\begin{figure}[h!]
    \centering
    \includegraphics[width=0.9\textwidth]{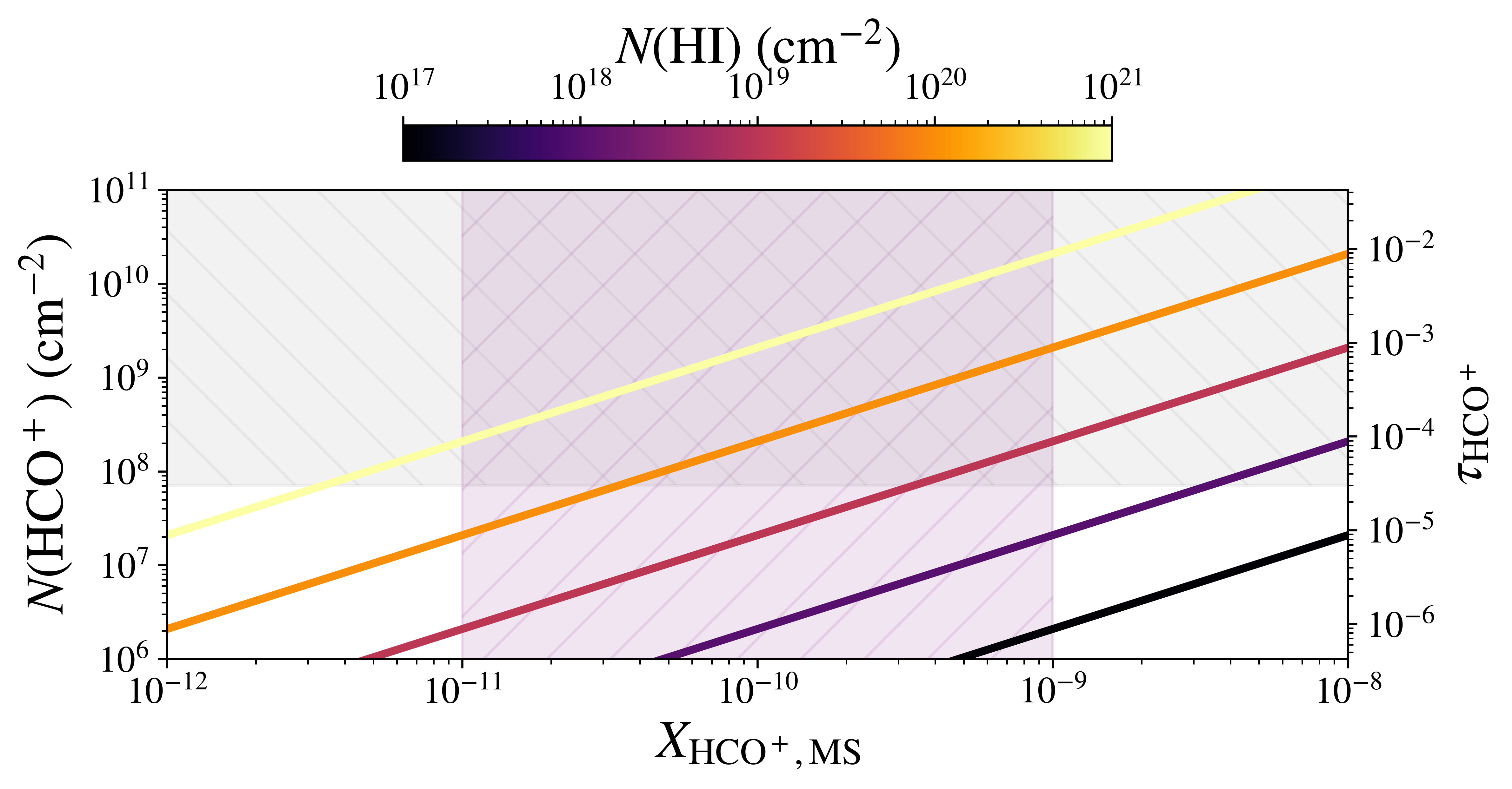}
    \caption{The \hcop{} column density (left $y$-axis) expected if $f_{\mathrm{mol}}=0.04$ in the MS \citep{Sembach2001,Wakker2006,Richter2013} as a function of $X_{\hcop{}, \mathrm{MS}}$. The right $y$-axis shows the corresponding \hcop{} optical depth (assuming a Gaussian line with $2~\kms{}$ FWHM). Lines show the results for $N(\mathrm{HI})=10^{17}~\persc{},10^{18}~\persc{},10^{19}~\persc{},10^{20}~\persc{},$ and $10^{21}~\persc{}$ (colors correspond to the HI column density). The region highlighted in purple outlines the range of plausible $X_{\hcop{}, \mathrm{MS}}$ (Section \ref{sec:fmol}). The region highlighted in gray indicates the range of $\tau_{\hcop{}}$ that could be detected with ALMA in $\lesssim15~\mathrm{hr}$ for the sources listed in Table \ref{tab_coordinates}.}
    \label{fig:fmol_0.1}
\end{figure}

In Figure \ref{fig:fmol_0.1}, we show the \hcop{} column density implied by $f_{\mathrm{mol}}=0.04$ \citep{Sembach2001,Wakker2006,Richter2013} as a function of $X_{\hcop{}, \mathrm{MS}}$. Lines of different color show the results for different HI column densities. On the right side of the plot, we show the \hcop{} optical depth (assuming a FWHM of $2~\kms{}$) corresponding to the \hcop{} column density on the left axis. As in Figure \ref{fig:fmol}, we highlight in purple the range of $X_{\hcop{}, \mathrm{MS}}$ that we deem most plausible for the MS. We further highlight in gray the range of \hcop{} optical depths that could plausibly be detected at a level $\geq3\sigma$ with ALMA for sources with $90~\mathrm{GHz}$ flux densities listed in Table \ref{tab_coordinates} (note that the flux density sensitivity needed to reach an optical depth sensitivity $\sigma_{\tau}$ is $\sigma_{F}=\sigma_{\tau}\times F$, where $F$ is the flux density of the background source in Jy). The intersection of the puple- and gray-highlighted regions in Figure \ref{fig:fmol_0.1} (highlighted with cross-hatching) represents the space in which we could detect \hcop{} with ALMA in $\lesssim15~\mathrm{hr}$ if $f_{\mathrm{mol}}=0.04$.
For any sightline with an HI column density $\gtrsim10^{20}~\persc{}$, ALMA should be able to detect \hcop{} if $f_{\mathrm{mol}}\sim0.04$. For $10^{19}~\persc{} \lesssim N(\mathrm{HI}) \lesssim 10^{20}~\persc{}$, we could reasonably expect to detect \hcop{} if $X_{\hcop{}, \mathrm{MS}}\gtrsim\times10^{-10}$. It is unlikely that \hcop{} could be detected for sightlines with $N(\mathrm{HI})\lesssim10^{19}~\persc{}$.

\section{Conclusions}
We have carried out the most sensitive search to date for molecular gas at millimeter wavelengths towards the MS. Despite our excellent sensitivity, we do not detect absorption from any of the four molecular species we observed --- \hcop{}, HCN, HNC, and C$_2$H. We still do not detect absorption in any of these species when we stack the absorption spectra from our 10 sightlines, reaching an optical depth sensitivity of $10^{-3}$ in each of the four lines (Table \ref{tab:stacked_spectra}). 
This includes our observations in the direction of J0454-8101 (toward the Magellanic Bridge), where \citet{Murray15} reported a tentative detection of \hcop{} absorption. 
Despite our improved optical depth sensitivity ($\sigma_{\tau}=0.009$ per $0.3~\kms{}$ channel, versus $\sigma_{\tau}=0.026$ per $0.8~\kms{}$ channel in \citealt{Murray15}); we are unable to reproduce their result.

Using complementary HI emission \citep{McClure_Griffiths09,Peek11,Peek18}, we look at the environments in the MS where we fail to detect molecular gas (Figures \ref{fig_stream} and \ref{fig:stars}) and place upper limits on the fraction of hydrogen in H$_2$ in these environments. Over a plausible range of \hcop{} abundances in the MS, we find that the column densities are so low that, even with our excellent sensitivity, our \hcop{} column density limits only constrain the molecular fraction as $\lesssim90$\% in most cases (if the \hcop{} abundance is $\gtrsim10^{-10}$; see Figure \ref{fig:fmol}). For the stacked spectrum, we can place slightly more stringent limits, with an upper limit of $f_{\mathrm{mol}}\sim70$--$95\%$ over a range of possible \hcop{} abundances.
Previous estimates of the molecular fraction in the MS, derived from direct observations of \htwo{} and other species in UV absorption, range from extremely low, $\sim10^{-5}$--$10^{-4}$ \citep{Lehner02}, to more moderate, $\sim\mathrm{few}\times10^{-2}$ \citep{Sembach2001,Wakker2006,Richter2013}.
To be sensitive to this latter range, we estimate that an \hcop{} optical depth sensitivity of $\mathcal{O}(10^{-5})$ to $\mathcal{O}(10^{-3})$ is required (note this depends on both the HI column density and \hcop{} abundance; see Equation \ref{eq:fmol}). 
For background sources as bright as those observed in this project, we show that ALMA would be capable of detecting such diffuse molecular gas in an observing time $\lesssim16~\mathrm{hr}$ \footnote{See the ALMA Sensitivity Calculator, \url{https://almascience.eso.org/proposing/sensitivity-calculator}.}  if the HI column density is $\gtrsim10^{19}~\persc{}$ (see Figure \ref{fig:fmol_0.1}). 
This suggests that ALMA offers an important way
to test the existence of molecular gas in the MS,
though this requires considerably better sensitivity than achieved here (note that our sources were observed for $\leq10~\mathrm{min}$ in ALMA Cycle 2).


\section*{Acknowledgements}

We note the passing of our scientific colleague Prof. Blair Savage on 2022 July 19 in Madison, Wisconsin. With his students, postdocs, and colleagues, Blair made seminal contributions to studies
of the interstellar and intergalactic medium. His enthusiasm for studying molecular hydrogen and dust in the Magellanic Stream was infectious. 
We would like to thank Nickolas Pingel and Trey Wenger for their insight on this project and continual discussion at group meetings. We would also like to thank Scott Lucchini and Tobias Westmeier for their help in understanding the Magellanic Stream Coordinate System.
This project was partially funded by a Hilldale Undergraduate Research Fellowship, awarded in 2021 by the University of Wisconsin--Madison. We would like to thank both the Undergraduate Research Scholars Program at the University of Wisconsin--Madison and the Hilldale Undergraduate Research Fellowship for helping to make this project possible.
This paper makes use of the following ALMA data: ADS/JAO.ALMA\#2013.1.00700.S. ALMA is a partnership of ESO (representing its member states), NSF (USA) and NINS (Japan), together with NRC (Canada), MOST and ASIAA (Taiwan), and KASI (Republic of Korea), in cooperation with the Republic of Chile. The Joint ALMA Observatory is operated by ESO, AUI/NRAO and NAOJ. The National Radio Astronomy Observatory is a facility of the National Science Foundation operated under cooperative agreement by Associated Universities, Inc.


\bibliographystyle{aa}
\bibliography{LibraryOfAlexandria}

\begin{thebibliography}{68}
\expandafter\ifx\csname natexlab\endcsname\relax\def\natexlab#1{#1}\fi

\bibitem[{{Barger} {et~al.}(2017){Barger}, {Madsen}, {Fox}, {Wakker}, {Bland-Hawthorn}, {Nidever}, {Haffner}, {Antwi-Danso}, {Hernandez}, {Lehner}, {Hill}, {Curzons}, \& {Tepper-Garc{\'\i}a}}]{Barger2017}
{Barger}, K.~A., {Madsen}, G.~J., {Fox}, A.~J., {et~al.} 2017, \apj, 851, 110

\bibitem[{{Besla} {et~al.}(2010){Besla}, {Kallivayalil}, {Hernquist}, {van der Marel}, {Cox}, \& {Kere{\v{s}}}}]{Besla10}
{Besla}, G., {Kallivayalil}, N., {Hernquist}, L., {et~al.} 2010, \apjl, 721, L97

\bibitem[{{Casetti-Dinescu} {et~al.}(2014){Casetti-Dinescu}, {Moni Bidin}, {Girard}, {M{\'e}ndez}, {Vieira}, {Korchagin}, \& {van Altena}}]{Casetti-Dinescu14}
{Casetti-Dinescu}, D.~I., {Moni Bidin}, C., {Girard}, T.~M., {et~al.} 2014, \apjl, 784, L37

\bibitem[{{Chandra} {et~al.}(2023){Chandra}, {Naidu}, {Conroy}, {Bonaca}, {Zaritsky}, {Cargile}, {Caldwell}, {Johnson}, {Han}, \& {Ting}}]{Chandra2023}
{Chandra}, V., {Naidu}, R.~P., {Conroy}, C., {et~al.} 2023, arXiv e-prints, arXiv:2306.15719

\bibitem[{{Dempsey} {et~al.}(2020){Dempsey}, {McClure-Griffiths}, {Jameson}, \& {Buckland-Willis}}]{Dempsey20}
{Dempsey}, J., {McClure-Griffiths}, N.~M., {Jameson}, K., \& {Buckland-Willis}, F. 2020, \mnras, 496, 913

\bibitem[{{D'Onghia} \& {Fox}(2016)}]{DonghiaFox2016}
{D'Onghia}, E. \& {Fox}, A.~J. 2016, \araa, 54, 363

\bibitem[{{Draine}(2011)}]{DraineTextbook}
{Draine}, B.~T. 2011, {Physics of the Interstellar and Intergalactic Medium}

\bibitem[{{For} {et~al.}(2014){For}, {Staveley-Smith}, {Matthews}, \& {McClure-Griffiths}}]{For14}
{For}, B.~Q., {Staveley-Smith}, L., {Matthews}, D., \& {McClure-Griffiths}, N.~M. 2014, \apj, 792, 43

\bibitem[{{Fox}(2005)}]{Fox2005}
{Fox}, A.~J. 2005, PhD thesis, University of Wisconsin, Madison

\bibitem[{{Fox} {et~al.}(2013){Fox}, {Richter}, {Wakker}, {Lehner}, {Howk}, \& {Bland-Hawthorn}}]{Fox13}
{Fox}, A.~J., {Richter}, P., {Wakker}, B.~P., {et~al.} 2013, The Messenger, 153, 28

\bibitem[{{Fox} {et~al.}(2014){Fox}, {Wakker}, {Barger}, {Hernandez}, {Richter}, {Lehner}, {Bland-Hawthorn}, {Charlton}, {Westmeier}, {Thom}, {Tumlinson}, {Misawa}, {Howk}, {Haffner}, {Ely}, {Rodriguez-Hidalgo}, \& {Kumari}}]{Fox2014}
{Fox}, A.~J., {Wakker}, B.~P., {Barger}, K.~A., {et~al.} 2014, \apj, 787, 147

\bibitem[{{Galametz} {et~al.}(2020){Galametz}, {Schruba}, {De Breuck}, {Immer}, {Chevance}, {Galliano}, {Gusdorf}, {Lebouteiller}, {Lee}, {Madden}, {Polles}, \& {van Kempen}}]{Galametz20}
{Galametz}, M., {Schruba}, A., {De Breuck}, C., {et~al.} 2020, \aap, 643, A63

\bibitem[{{Gerin} {et~al.}(2019){Gerin}, {Liszt}, {Neufeld}, {Godard}, {Sonnentrucker}, {Pety}, \& {Roueff}}]{Gerin2019}
{Gerin}, M., {Liszt}, H., {Neufeld}, D., {et~al.} 2019, \aap, 622, A26

\bibitem[{{Godard} {et~al.}(2010){Godard}, {Falgarone}, {Gerin}, {Hily-Blant}, \& {de Luca}}]{Godard2010}
{Godard}, B., {Falgarone}, E., {Gerin}, M., {Hily-Blant}, P., \& {de Luca}, M. 2010, \aap, 520, A20

\bibitem[{{Gong} {et~al.}(2017){Gong}, {Ostriker}, \& {Wolfire}}]{Gong2017}
{Gong}, M., {Ostriker}, E.~C., \& {Wolfire}, M.~G. 2017, \apj, 843, 38

\bibitem[{{Hafner} {et~al.}(2023){Hafner}, {Dawson}, {Nguyen}, {Heiles}, {Wardle}, {Lee}, {Murray}, {Thompson}, \& {Stanimirovi{\'c}}}]{Hafner23}
{Hafner}, A., {Dawson}, J.~R., {Nguyen}, H., {et~al.} 2023, \pasa, 40, e015

\bibitem[{{HI4PI Collaboration} {et~al.}(2016){HI4PI Collaboration}, {Ben Bekhti}, {Fl{\"o}er}, {Keller}, {Kerp}, {Lenz}, {Winkel}, {Bailin}, {Calabretta}, {Dedes}, {Ford}, {Gibson}, {Haud}, {Janowiecki}, {Kalberla}, {Lockman}, {McClure-Griffiths}, {Murphy}, {Nakanishi}, {Pisano}, \& {Staveley-Smith}}]{HI4PI}
{HI4PI Collaboration}, {Ben Bekhti}, N., {Fl{\"o}er}, L., {et~al.} 2016, \aap, 594, A116

\bibitem[{{Irwin} {et~al.}(1985){Irwin}, {Kunkel}, \& {Demers}}]{Irwin85}
{Irwin}, M.~J., {Kunkel}, W.~E., \& {Demers}, S. 1985, \nat, 318, 160

\bibitem[{{Keller} \& {Wood}(2006)}]{Keller06}
{Keller}, S.~C. \& {Wood}, P.~R. 2006, \apj, 642, 834

\bibitem[{{Kim} {et~al.}(2024){Kim}, {Zheng}, \& {Putman}}]{Kim2024}
{Kim}, D.~A., {Zheng}, Y., \& {Putman}, M.~E. 2024, arXiv e-prints, arXiv:2402.08810

\bibitem[{{Kosenko} \& {Balashev}(2023)}]{Kosenko2023}
{Kosenko}, D.~N. \& {Balashev}, S.~A. 2023, \mnras, 525, 2820

\bibitem[{{Krishnarao} {et~al.}(2022){Krishnarao}, {Fox}, {D'Onghia}, {Wakker}, {Cashman}, {Howk}, {Lucchini}, {French}, \& {Lehner}}]{Krishnarao22}
{Krishnarao}, D., {Fox}, A.~J., {D'Onghia}, E., {et~al.} 2022, \nat, 609, 915

\bibitem[{{Larsson} {et~al.}(2008){Larsson}, {McCall}, \& {Orel}}]{Larsson2008}
{Larsson}, M., {McCall}, B.~J., \& {Orel}, A.~E. 2008, Chemical Physics Letters, 462, 145

\bibitem[{{Lehner}(2002)}]{Lehner02}
{Lehner}, N. 2002, \apj, 578, 126

\bibitem[{{Liszt} \& {Gerin}(2023)}]{LisztGerin2023}
{Liszt}, H. \& {Gerin}, M. 2023, \apj, 943, 172

\bibitem[{{Liszt} \& {Lucas}(2000)}]{LisztLucas2000}
{Liszt}, H. \& {Lucas}, R. 2000, \aap, 355, 333

\bibitem[{{Liszt} \& {Lucas}(2001)}]{LisztLucas2001_HCN_HNC}
{Liszt}, H. \& {Lucas}, R. 2001, \aap, 370, 576

\bibitem[{{Lucas} \& {Liszt}(1996)}]{LucasLiszt1996}
{Lucas}, R. \& {Liszt}, H. 1996, \aap, 307, 237

\bibitem[{{Lucas} \& {Liszt}(2000)}]{LucasLiszt2000_CCH}
{Lucas}, R. \& {Liszt}, H.~S. 2000, \aap, 358, 1069

\bibitem[{{Lucchini} {et~al.}(2021){Lucchini}, {D'Onghia}, \& {Fox}}]{Lucchini21}
{Lucchini}, S., {D'Onghia}, E., \& {Fox}, A.~J. 2021, \apjl, 921, L36

\bibitem[{{Luo} {et~al.}(2020){Luo}, {Li}, {Tang}, {Dawson}, {Dickey}, {Bronfman}, {Qin}, {Gibson}, {Plambeck}, {Finger}, {Green}, {Mardones}, {Koo}, \& {Lo}}]{Luo2020}
{Luo}, G., {Li}, D., {Tang}, N., {et~al.} 2020, \apjl, 889, L4

\bibitem[{{Mac Low} \& {Klessen}(2004)}]{MacLowKlessen2004}
{Mac Low}, M.-M. \& {Klessen}, R.~S. 2004, Reviews of Modern Physics, 76, 125

\bibitem[{{Mathewson} {et~al.}(1979){Mathewson}, {Ford}, {Schwarz}, \& {Murray}}]{Mathewson79}
{Mathewson}, D.~S., {Ford}, V.~L., {Schwarz}, M.~P., \& {Murray}, J.~D. 1979, in The Large-Scale Characteristics of the Galaxy, ed. W.~B. {Burton}, Vol.~84, 547

\bibitem[{{Matthews} {et~al.}(2009){Matthews}, {Kirk}, {Johnstone}, {Weferling}, {Cohen}, {Jenness}, {Evans}, {Davis}, {Dent}, {Fuller}, {Jackson}, {Rathborne}, {Richer}, \& {Simon}}]{Matthews2009}
{Matthews}, H., {Kirk}, H., {Johnstone}, D., {et~al.} 2009, \aj, 138, 1380

\bibitem[{{McClure-Griffiths} {et~al.}(2009){McClure-Griffiths}, {Pisano}, {Calabretta}, {Ford}, {Lockman}, {Staveley-Smith}, {Kalberla}, {Bailin}, {Dedes}, {Janowiecki}, {Gibson}, {Murphy}, {Nakanishi}, \& {Newton-McGee}}]{McClure_Griffiths09}
{McClure-Griffiths}, N.~M., {Pisano}, D.~J., {Calabretta}, M.~R., {et~al.} 2009, \apjs, 181, 398

\bibitem[{{McClure-Griffiths} {et~al.}(2008){McClure-Griffiths}, {Staveley-Smith}, {Lockman}, {Calabretta}, {Ford}, {Kalberla}, {Murphy}, {Nakanishi}, \& {Pisano}}]{McClure-Griffiths08}
{McClure-Griffiths}, N.~M., {Staveley-Smith}, L., {Lockman}, F.~J., {et~al.} 2008, \apjl, 673, L143

\bibitem[{{Millar} \& {Herbst}(1990)}]{MillarHerbst1990}
{Millar}, T.~J. \& {Herbst}, E. 1990, \mnras, 242, 92

\bibitem[{{Mizuno} {et~al.}(2006){Mizuno}, {Muller}, {Maeda}, {Kawamura}, {Minamidani}, {Onishi}, {Mizuno}, \& {Fukui}}]{Mizuno2006}
{Mizuno}, N., {Muller}, E., {Maeda}, H., {et~al.} 2006, \apjl, 643, L107

\bibitem[{{Muller} {et~al.}(2003){Muller}, {Staveley-Smith}, \& {Zealey}}]{Muller_2003}
{Muller}, E., {Staveley-Smith}, L., \& {Zealey}, W.~J. 2003, \mnras, 338, 609

\bibitem[{{Murray} {et~al.}(2018){Murray}, {Stanimirovi{\'c}}, {Goss}, {Heiles}, {Dickey}, {Babler}, \& {Kim}}]{Murray18}
{Murray}, C.~E., {Stanimirovi{\'c}}, S., {Goss}, W.~M., {et~al.} 2018, \apjs, 238, 14

\bibitem[{{Murray} {et~al.}(2015){Murray}, {Stanimirovi{\'c}}, {McClure-Griffiths}, {Putman}, {Liszt}, {Wong}, {Richter}, {Dawson}, {Dickey}, {Lindner}, {Babler}, \& {Allison}}]{Murray15}
{Murray}, C.~E., {Stanimirovi{\'c}}, S., {McClure-Griffiths}, N.~M., {et~al.} 2015, \apj, 808, 41

\bibitem[{{Nguyen} {et~al.}(2019){Nguyen}, {Dawson}, {Lee}, {Murray}, {Stanimirovi{\'c}}, {Heiles}, {Miville-Desch{\^e}nes}, \& {Petzler}}]{Nguyen19}
{Nguyen}, H., {Dawson}, J.~R., {Lee}, M.-Y., {et~al.} 2019, \apj, 880, 141

\bibitem[{{Nidever} {et~al.}(2008){Nidever}, {Majewski}, \& {Butler Burton}}]{Nidever08}
{Nidever}, D.~L., {Majewski}, S.~R., \& {Butler Burton}, W. 2008, \apj, 679, 432

\bibitem[{{Nidever} {et~al.}(2010){Nidever}, {Majewski}, {Butler Burton}, \& {Nigra}}]{Nidever10}
{Nidever}, D.~L., {Majewski}, S.~R., {Butler Burton}, W., \& {Nigra}, L. 2010, \apj, 723, 1618

\bibitem[{{Panessa} {et~al.}(2023){Panessa}, {Seifried}, {Walch}, {Gaches}, {Barnes}, {Bigiel}, \& {Neumann}}]{Panessa2023}
{Panessa}, M., {Seifried}, D., {Walch}, S., {et~al.} 2023, \mnras, 523, 6138

\bibitem[{{Park} {et~al.}(2023){Park}, {Lee}, {Bialy}, {Burkhart}, {Dawson}, {Heiles}, {Li}, {Murray}, {Nguyen}, {Petzler}, \& {Stanimirovi{\'c}}}]{Park23}
{Park}, G., {Lee}, M.-Y., {Bialy}, S., {et~al.} 2023, in IAU Symposium, Vol. 373, Resolving the Rise and Fall of Star Formation in Galaxies, ed. T.~{Wong} \& W.-T. {Kim}, 81--86

\bibitem[{{Peek} {et~al.}(2018){Peek}, {Babler}, {Zheng}, {Clark}, {Douglas}, {Korpela}, {Putman}, {Stanimirovic}, {Gibson}, \& {Heiles}}]{Peek18}
{Peek}, J.~E.~G., {Babler}, B.~L., {Zheng}, Y., {et~al.} 2018, VizieR Online Data Catalog, J/ApJS/234/2

\bibitem[{{Peek} {et~al.}(2011){Peek}, {Heiles}, {Douglas}, {Lee}, {Grcevich}, {Stanimirovi{\'c}}, {Putman}, {Korpela}, {Gibson}, {Begum}, {Saul}, {Robishaw}, \& {Kr{\v{c}}o}}]{Peek11}
{Peek}, J.~E.~G., {Heiles}, C., {Douglas}, K.~A., {et~al.} 2011, \apjs, 194, 20

\bibitem[{{Price-Whelan} {et~al.}(2019){Price-Whelan}, {Nidever}, {Choi}, {Schlafly}, {Morton}, {Koposov}, \& {Belokurov}}]{Price-Whelan19}
{Price-Whelan}, A.~M., {Nidever}, D.~L., {Choi}, Y., {et~al.} 2019, \apj, 887, 19

\bibitem[{{Putman} {et~al.}(2003){Putman}, {Staveley-Smith}, {Freeman}, {Gibson}, \& {Barnes}}]{Putman03}
{Putman}, M.~E., {Staveley-Smith}, L., {Freeman}, K.~C., {Gibson}, B.~K., \& {Barnes}, D.~G. 2003, \apj, 586, 170

\bibitem[{{Richter} {et~al.}(2018){Richter}, {Fox}, {Wakker}, {Howk}, {Lehner}, {Barger}, {D'Onghia}, \& {Lockman}}]{Richter18}
{Richter}, P., {Fox}, A.~J., {Wakker}, B.~P., {et~al.} 2018, \apj, 865, 145

\bibitem[{{Richter} {et~al.}(2013){Richter}, {Fox}, {Wakker}, {Lehner}, {Howk}, {Bland-Hawthorn}, {Ben Bekhti}, \& {Fechner}}]{Richter2013}
{Richter}, P., {Fox}, A.~J., {Wakker}, B.~P., {et~al.} 2013, \apj, 772, 111

\bibitem[{{Richter} {et~al.}(2001){Richter}, {Sembach}, {Wakker}, \& {Savage}}]{Richter2001}
{Richter}, P., {Sembach}, K.~R., {Wakker}, B.~P., \& {Savage}, B.~D. 2001, \apjl, 562, L181

\bibitem[{{Rolleston} {et~al.}(2002){Rolleston}, {Trundle}, \& {Dufton}}]{Rolleston02}
{Rolleston}, W.~R.~J., {Trundle}, C., \& {Dufton}, P.~L. 2002, \aap, 396, 53

\bibitem[{{Roman-Duval} {et~al.}(2014){Roman-Duval}, {Gordon}, {Meixner}, {Bot}, {Bolatto}, {Hughes}, {Wong}, {Babler}, {Bernard}, {Clayton}, {Fukui}, {Galametz}, {Galliano}, {Glover}, {Hony}, {Israel}, {Jameson}, {Lebouteiller}, {Lee}, {Li}, {Madden}, {Misselt}, {Montiel}, {Okumura}, {Onishi}, {Panuzzo}, {Reach}, {Remy-Ruyer}, {Robitaille}, {Rubio}, {Sauvage}, {Seale}, {Sewilo}, {Staveley-Smith}, \& {Zhukovska}}]{Roman-Duval14}
{Roman-Duval}, J., {Gordon}, K.~D., {Meixner}, M., {et~al.} 2014, \apj, 797, 86

\bibitem[{{Rybarczyk} {et~al.}(2022{\natexlab{a}}){Rybarczyk}, {Gong}, {Stanimirovi{\'c}}, {Babler}, {Murray}, {Winters}, {Luo}, {Dame}, \& {Steffes}}]{Rybarczyk22a}
{Rybarczyk}, D.~R., {Gong}, M., {Stanimirovi{\'c}}, S., {et~al.} 2022{\natexlab{a}}, \apj, 926, 190

\bibitem[{{Rybarczyk} {et~al.}(2022{\natexlab{b}}){Rybarczyk}, {Stanimirovi{\'c}}, {Gong}, {Babler}, {Murray}, {Gerin}, {Winters}, {Luo}, {Dame}, \& {Steffes}}]{Rybarczyk22b}
{Rybarczyk}, D.~R., {Stanimirovi{\'c}}, S., {Gong}, M., {et~al.} 2022{\natexlab{b}}, \apj, 928, 79

\bibitem[{{Sembach} {et~al.}(2001){Sembach}, {Howk}, {Savage}, \& {Shull}}]{Sembach2001}
{Sembach}, K.~R., {Howk}, J.~C., {Savage}, B.~D., \& {Shull}, J.~M. 2001, \aj, 121, 992

\bibitem[{{Stanimirovic} {et~al.}(2010){Stanimirovic}, {Gallagher}, \& {Nigra}}]{Stanimirovic10}
{Stanimirovic}, S., {Gallagher}, J.~S., I., \& {Nigra}, L. 2010, Serbian Astronomical Journal, 180, 1

\bibitem[{{Stanimirovi{\'c}} {et~al.}(2008){Stanimirovi{\'c}}, {Hoffman}, {Heiles}, {Douglas}, {Putman}, \& {Peek}}]{Stanimirovic08}
{Stanimirovi{\'c}}, S., {Hoffman}, S., {Heiles}, C., {et~al.} 2008, \apj, 680, 276

\bibitem[{{Stanimirovi{\'c}} {et~al.}(2014){Stanimirovi{\'c}}, {Murray}, {Lee}, {Heiles}, \& {Miller}}]{Stanimirovic14}
{Stanimirovi{\'c}}, S., {Murray}, C.~E., {Lee}, M.-Y., {Heiles}, C., \& {Miller}, J. 2014, \apj, 793, 132

\bibitem[{{Sternberg} {et~al.}(2002){Sternberg}, {McKee}, \& {Wolfire}}]{Sternberg2002}
{Sternberg}, A., {McKee}, C.~F., \& {Wolfire}, M.~G. 2002, \apjs, 143, 419

\bibitem[{{Wakker}(2001)}]{Wakker01}
{Wakker}, B.~P. 2001, \apjs, 136, 463

\bibitem[{{Wakker}(2006)}]{Wakker2006}
{Wakker}, B.~P. 2006, \apjs, 163, 282

\bibitem[{{Weiner} \& {Williams}(1996)}]{Weiner1996}
{Weiner}, B.~J. \& {Williams}, T.~B. 1996, \aj, 111, 1156

\bibitem[{{Westmeier}(2018)}]{Westmeier18}
{Westmeier}, T. 2018, \mnras, 474, 289

\bibitem[{{Wolfire} {et~al.}(1995){Wolfire}, {McKee}, {Hollenbach}, \& {Tielens}}]{Wolfire1995}
{Wolfire}, M.~G., {McKee}, C.~F., {Hollenbach}, D., \& {Tielens}, A.~G.~G.~M. 1995, \apj, 453, 673

\bibitem[{{Zaritsky} {et~al.}(2020){Zaritsky}, {Conroy}, {Naidu}, {Cargile}, {Putman}, {Besla}, {Bonaca}, {Caldwell}, {Han}, {Johnson}, {Speagle}, \& {Ting}}]{Zaritsky20}
{Zaritsky}, D., {Conroy}, C., {Naidu}, R.~P., {et~al.} 2020, \apjl, 905, L3

\end{thebibliography}

\end{document}